\documentclass{PoS}
\usepackage{graphicx}
\usepackage{slashed}
\usepackage{psfrag}

\PoS{PoS(LAT2005)267}
\title{Monte-Carlo simulation of the chiral Gross-Neveu model\footnote{preprint: HU-EP-05/59, DESY 05-188, SFB/CPP-05-61}}
\ShortTitle{MC simulation of the chiral GN-model}
\author{\speaker{Tomasz Korzec}, Francesco Knechtli, Ulli Wolff\\
        Humboldt Universit\"at zu Berlin, Institut f\"ur Physik\\
	Newtonstrasse 15, 12489 Berlin, Germany\\
	E-mail: \email{\{korzec,knechtli,uwolff\}@physik.hu-berlin.de}}
\author{Bj\"orn Leder\\
        DESY, Platanenallee 6, 15738 Zeuthen, Germany\\
	E-mail: \email{bjoern.leder@desy-zeuthen.de}}	
	
\abstract{We investigate the two flavor chiral Gross-Neveu model in the Schr\"odinger functional on the lattice.
The procedure necessary to recover
chiral symmetry in the continuum limit of this model with Wilson fermions
is discussed. We introduce several useful observables and present 
a first demonstration of the feasibility  of Monte-Carlo simulations in this model.}
\FullConference{XXIIIrd International Symposium on Lattice Field Theory\\

                 25-30 July 2005\\

                 Trinity College, Dublin, Ireland}

\begin{document}
\section{Introduction}
   The Monte-Carlo simulation of lattice QCD with dynamical fermions is rather expensive.  
   Therefore ideas and concepts are sometimes tested in low dimensional toy-models which are comparatively
   cheap and often provide interesting insights. 
   Two of these toy models are the two dimensional Gross-Neveu and the chiral Gross-Neveu model. They are
   renormalizable, asymptotically free and have a rich particle spectrum. The masses of the particles 
   are generated dynamically. The chiral model is invariant under axial $U(1)$ transformations, the 
   standard Gross-Neveu model only under a discrete $Z_2$-subgroup. Both models were treated in the large-$N$ 
   approximation~\cite{Gross1974,Aoki1985,Knechtli2004}
   and in perturbation theory~\cite{Luperini1991,Kenna:2001fs} in the past and as both models are integrable also some exact 
   results are available~\cite{Karowski:1980kq,Forgacs1991}. Different discretizations were used to simulate the discrete 
   model~\cite{Cohen1981,Campostrini1987}. 
   To our knowledge the chiral model has so far only
   been simulated with staggered fermions \cite{Hands1992} in 3d. Here we present first results of simulations 
   with Wilson fermions with which the restoration of chiral symmetry requires more care \cite{Bochicchio:1985xa}.
   
\section{The chiral Gross-Neveu model}
   The chiral Gross-Neveu model describes $N$ flavors of self-interacting fermions in two dimensions.
   The standard way of writing its euclidian action is
   \begin{equation}\label{standardaction}
      S = \int \! d^2x \ \left\{ \bar\psi \slashed\partial \psi - \frac{g_S^2}{2} \left[ (\bar\psi \psi)^2 + (\bar\psi i\gamma_5 \psi)^2 \right]
      -\frac{g_V^2}{2} \bar\psi\gamma_\mu \psi\bar\psi\gamma_\mu \psi  \right\} \, ,
   \end{equation}
   where the summations in flavor space are left implicit.
   In two dimensions such quartic interaction terms are renormalizable. Moreover the model is asymptotically free
   which makes it in a sense similar to QCD.
   
   An equivalent action is given by
   \begin{equation}
      S = \int \! d^2x \ \left\{\bar\psi[\slashed\partial + \sigma + i\gamma_5\Pi + \slashed A]\psi + 
      \frac{\sigma^2 + \Pi^2}{2g_S^2} + \frac{A_\mu A_\mu}{2g_V^2}\right\}\, .
   \end{equation}
   Here real auxiliary fields $\sigma$, $\Pi$ and $A_\mu$ have been introduced which are additional bosonic integration variables in 
   the path integral. If these are integrated out
   the action (\ref{standardaction}) is recovered, hence both formulations are equivalent on the level of the generating 
   functionals for fermionic correlation functions. The language with auxiliary fields is better accessible to numerical simulations because the fermionic
   fields enter bilinearly and can be integrated out.

\section{Lattice formulation with Wilson fermions}    
   On the lattice the action of the model with Wilson fermions takes the form
    \begin{equation}\label{latticeaction}
        S = a^2\sum_x \left\{\bar\psi[D_W
        +m_0+\sigma+i\gamma_5\Pi+\slashed A]\psi +\frac{\sigma^2}{2g_S^2}
        +\frac{\Pi^2}{2g_P^2}+\frac{A_\mu A_\mu}{2g_V^2} \right\} \, .
    \end{equation}
    The Wilson operator is given by
    \begin{equation}
       D_W = \frac{1}{2}\left[ \gamma_\mu(\nabla_\mu^*+\nabla_\mu)- a \nabla_\mu^*\nabla_\mu
     \right]\, ,
    \end{equation}
    with the forward and backward lattice differences
    \begin{eqnarray}
       \nabla_\mu \psi(x)  &=& \frac{1}{a}\left[e^{ia\theta_\mu/L}\psi(x+\hat\mu) - \psi(x)\right] \\
       \nabla^*_\mu\psi(x) &=& \frac{1}{a}\left[\psi(x) - e^{-ia\theta_\mu/L} \psi(x-\hat\mu)\right] \, .
    \end{eqnarray}
    We incorporate a phase factor into the definition which is equivalent to a particular choice of boundary 
    conditions. For instance $\theta=0$ corresponds to periodic and $\theta=\pi$ to antiperiodic boundaries.
    The fermionic matrix $D = D_W + m_0 + \sigma + i\gamma_5\Pi + \slashed A$ is neither hermitian nor anti-hermitian
    and unlike in QCD not even $\gamma_5$-hermitian. For periodic and antiperiodic boundaries however it is 
    real (in a Majorana representation of the $\gamma$-matrices).
    
    We simulate the model with a Hybrid Monte-Carlo algorithm \cite{Duane1987}. For this we introduce one complex 
    pseudo fermion field $\phi$ for each two flavors in order to represent the fermionic determinant
    as well as for each auxiliary field a conjugate momentum field. In the simulation of the 
    two flavor theory most processor time is used for the integration of the equations of motion (leap frog integrator)
    and in particular in the solution of linear systems $(DD^\dagger) x = \phi$ 
    which is done with the conjugate gradient method.

\section{The Schr\"odinger functional}
In computer simulations finite volume is unavoidable. We can exploit it by using finite volume renormalization
schemes,  a particularly 
successful one in QCD being the Schr\"odinger functional scheme \cite{Luscher:1992an,Sint:1993un}.
We simulate the chiral Gross-Neveu model in a finite box with spatial extent $L$ and temporal extent $T$. In the
spatial dimension we apply (anti-)periodic boundary conditions and in the temporal one we have Dirichlet
boundaries
\begin{eqnarray}
   \bar \psi(0,x_1)P_-   = \bar\rho(x_1)  &\quad& \ P_+\psi(0,x_1) = \rho(x_1)  \nonumber \\
   \bar \psi(T,x_1)P_+   = \bar\rho'(x_1) &\quad& \ P_-\psi(T,x_1) = \rho'(x_1) \, ,
\end{eqnarray}
where $P_\pm = (1 \pm \gamma_0)/2$.
Such boundary conditions may in principle cause new divergences, which would require the introduction
of boundary-counterterms into the action, but no such terms are necessary in our case \cite{Leder2005}.

An observable is a polynomial in the functional derivatives with respect to sources $\bar \eta$, $\eta$ and
boundary fields
\begin{eqnarray}
   \bar\psi(x)    \equiv -\frac{\delta}{\delta \eta(x)}     &\quad& \psi(x)    \equiv \frac{\delta}{\delta \bar\eta(x)} \\
   \bar\zeta(x_1) \equiv -\frac{\delta}{\delta \rho(x_1)}   &\quad& \zeta(x_1) \equiv \frac{\delta}{\delta \bar\rho(x_1)} \\
   \bar\zeta'(x_1)\equiv -\frac{\delta}{\delta \rho'(x_1)}  &\quad& \zeta'(x_1)\equiv \frac{\delta}{\delta \bar\rho'(x_1)}\, ,   
\end{eqnarray}
which acts on the generating functional.

\section{Chirally symmetric continuum limit}
In the formal continuum model a Ward identity associated with the axial $U(1)$ symmetry can be
derived
\begin{equation}\label{wardidentity}
   \frac{\partial}{\partial x_\mu} \langle A_\mu(x) {\mathcal O}(y) \rangle = 0\, , \quad \qquad x\neq y\, ,
\end{equation}
where $\mathcal O$ is some arbitrary operator.

In our lattice model chiral symmetry is broken explicitly by the Wilson term and we have to introduce
a bare mass $m_0 \neq 0$ and non-degenerate couplings $g^2_P \neq g^2_S$ into the action (\ref{latticeaction}) 
which increases the number of bare parameters to four: $m_0$, $g_S^2$, $g_P^2$
and $g_V^2$. Let us assume for a moment that we can define two different renormalized finite volume observables $\bar g_1$ and $\bar g_2$ and
that we know two operators ${\mathcal O}_1$ and ${\mathcal O}_2$, which make (\ref{wardidentity}) vanish only
due to chiral symmetry and not due to some other symmetry of the action (e.g. parity or charge conjugation). 
We want to take a continuum limit of some other observables e.g. the step-scaling functions of $\bar g_1$ and $\bar g_2$
at a fixed physical size of the lattice.
We then need to tune the four bare parameters for a series of increasingly larger lattices  in
such a way that $\bar g_1$ and $\bar g_2$ are kept at some fixed  values and at the same time (\ref{wardidentity}) is 
imposed for both operators ${\mathcal O}_1$ and ${\mathcal O}_2$. The continuum-extrapolated values
of the observables would be universal predictions of the chirally invariant theory.

At first sight the tuning of four parameters seems impracticable even in a two dimensional system. But there is
a formal argumentation in the continuum~\cite{Furuya:1982fh} that might help us a lot. There it is established
that the combination $g_V^2 - g_S^2/N$ does not renormalize and hence can be set to an arbitrary 
constant\footnote{We believe that this is the reason why the vector-vector
interaction term in the action is left out in most large-$N$ calculations.}
\begin{equation}\label{formalresult}
   g_V^2  = {\rm const} + \frac{g_S^2}{N} \, .
\end{equation} 
Although we do not see how the formal argumentation can be reproduced on the lattice one may hope that
(\ref{formalresult}) holds also there up to cutoff effects. 
For the tuning of $g_P$ and $m_0$ we can use large-$N$ results~\cite{Aoki1985} and perturbation 
theory~\cite{Leder2005} as a first guess.

\section{Observables}
\begin{figure}
   \psfrag{label1}{$g_S^2$}
   \psfrag{label2}{$am$}
   \psfrag{label5}{\footnotesize MC-data}
   \psfrag{label4}{\footnotesize Free theory}
   \psfrag{label3}{\footnotesize 1 loop PT}
   \includegraphics[width = 0.52\linewidth]{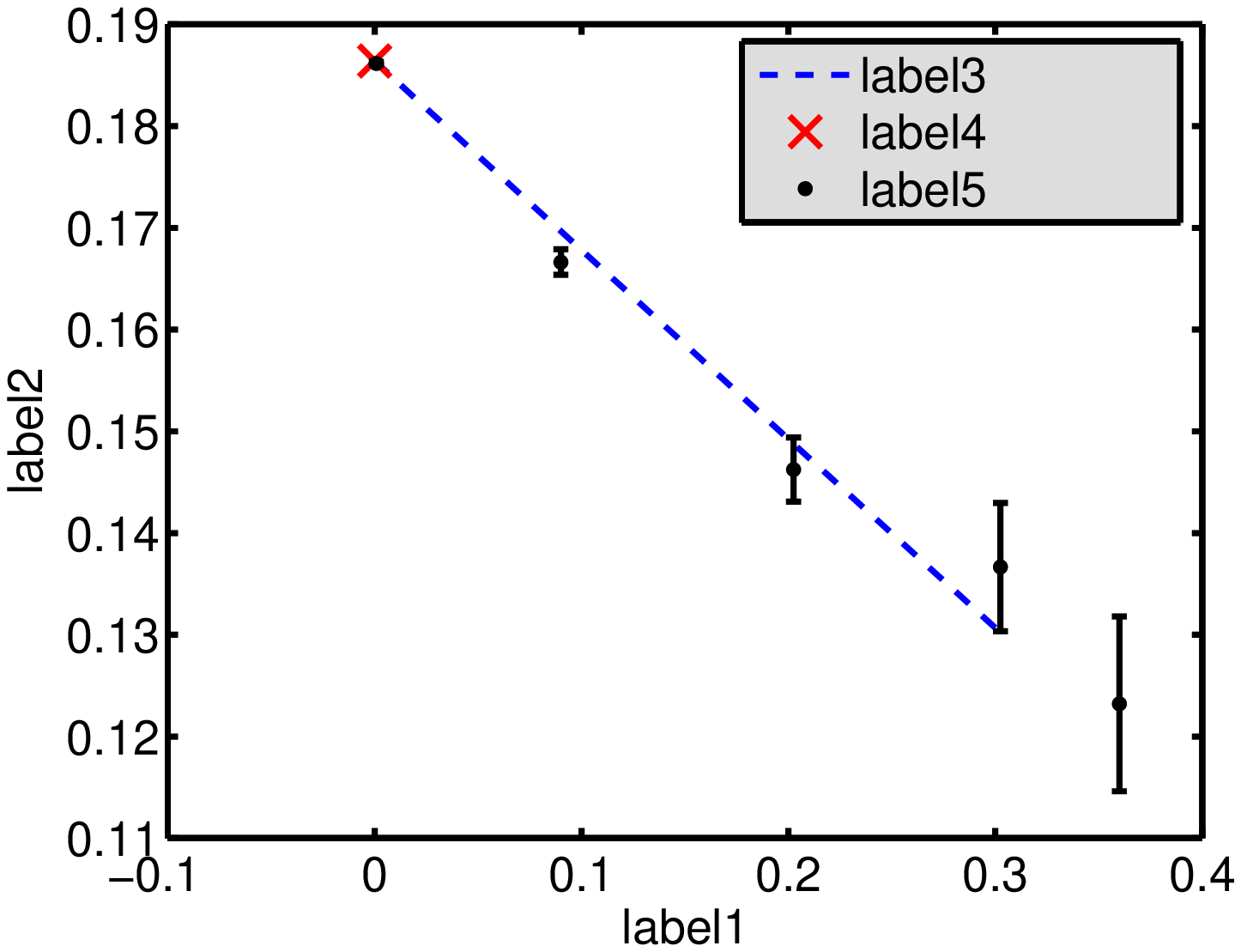} 
   \psfrag{label1}{$L/a$}
   \psfrag{label2}{$\partial^2 \log Z /\partial \theta^2 |_{\theta = 0}$}
   \psfrag{label3}{\footnotesize MC-data}
   \psfrag{label4}{\footnotesize Free theory}
   \includegraphics[width = 0.52\linewidth]{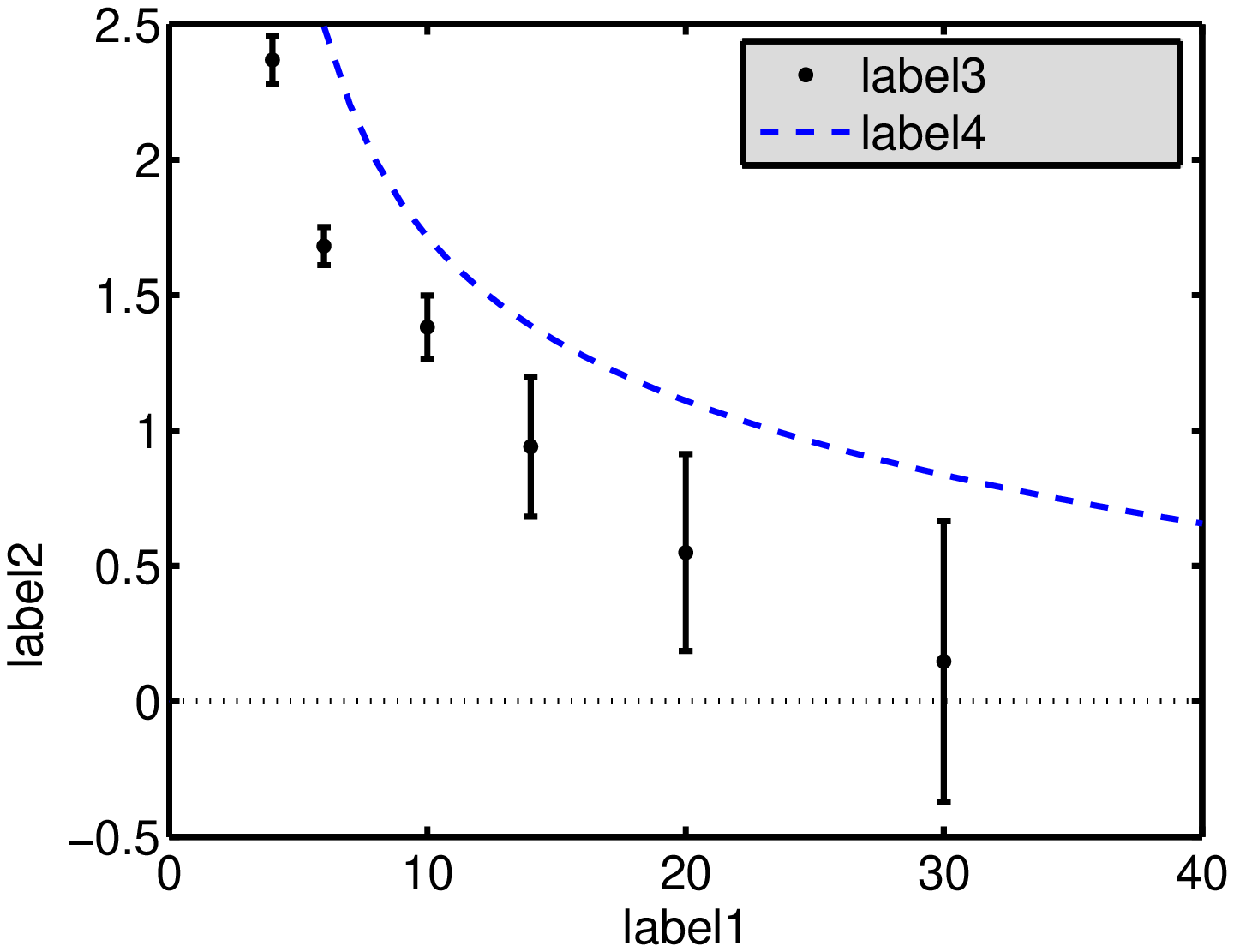}
   \caption{Left: The quotient $m$ versus the bare coupling $g_S^2$ on a lattice of size $L \times T =12 \times 13$.
                  The bare mass was set to $m_0 = m_c^{(1)} + 0.2$. The other couplings were $g_P = 10g_V = g_S$. 
            Right: The second derivative of the logarithm of the Schr\"odinger functional with respect to $\theta$ versus the
	          lattice size. The couplings and  the bare mass are kept constant at $m_0 = 0.023$, $g_S = g_P = g_V = 0.1$.}\label{allfigs}
\end{figure}

In our simulations we measure the following correlation functions
\begin{eqnarray}
   f_A(x_0) &=& -\frac{a^3}{L} \sum_{x_1,y_1,z_1} \left\langle \bar \psi(x)\gamma_0\gamma_5 \psi(x) \bar \zeta(y_1) \gamma_5 \zeta(z_1) \right\rangle\\
   f_P(x_0) &=& -\frac{a^3}{L} \sum_{x_1,y_1,z_1} \left\langle \bar \psi(x)\gamma_5 \psi(x) \bar \zeta(y_1) \gamma_5 \zeta(z_1) \right\rangle\\
   f_2 &=& -\frac{a^2}{N L} \sum_{y_1,y_1'} \left\langle \bar \zeta'(y_1') \zeta(y_1) \right\rangle \\
   f_4 &=& -\frac{a^4}{NL^2} \sum_{y_1,z_1,y_1',z_1'} \left\langle \bar \zeta'(y_1')\gamma_5\lambda^a\zeta'(z_1') \bar\zeta(y_1)\gamma_5\lambda^a\zeta(z_1)\right\rangle \, .
\end{eqnarray}
The quotient
\begin{equation}\label{mpcac}
   m(x_0) = \frac{\tilde \partial_0 f_A(x_0)}{2f_P(x_0)} 
\end{equation}
is independent of $x_0$ up to lattice artifacts and vanishes when the Ward identity (\ref{wardidentity}) is satisfied. On the left hand side of
figure \ref{allfigs} this quotient is plotted and compared with perturbation theory. The bare mass was set to 
$m_0 = m_c^{(1)}+0.2$, where $m_c^{(1)}$ is the 1-loop result for the critical mass.

We can use $f_2$ and $f_4$ to define a finite volume running coupling
\begin{equation}
   \bar g^2 \propto (f_4)_R - f_4^{(0)}\, , \qquad \quad (f_4)_R = Z_\zeta^4 f_4 = \frac{f_4}{(f_2)^2} (f_2^{(0)})^2 \, ,
\end{equation}
where $f_2^{(0)}$ and $f_4^{(0)}$ are the tree-level values. The last equality holds if we renormalize $f_2$ by requiring
that it takes its tree-level value
\begin{equation}
   (f_2)_R = Z_\zeta^2 f_2 \equiv f_2^{(0)}\, .
\end{equation}

Other renormalized observables can be obtained from derivatives of the logarithm of the Schr\"odinger-Functional
with respect to the angle $\theta$
\begin{eqnarray}
   \frac{\partial}{\partial \theta} \log Z \Bigr|_{\theta = 0} &=& 0 \\
   \frac{\partial^2}{\partial \theta^2} \log Z \Biggr|_{\theta = 0} 
   &=& \left\langle \frac{-a^3}{L^2}\sum_x \bar \psi(x)[1-\gamma_1]\psi(x+\hat 1) \right\rangle \nonumber \\
   &+& \biggl\langle \frac{-a^4}{2L^2}\sum_{x,y}\Bigl\{ \bar\psi(x)[1-\gamma_1]\psi(x+\hat 1) \bar\psi(y)[1-\gamma_1]\psi(y+\hat 1) \nonumber \\
   &-&                                          \bar\psi(x)[1-\gamma_1]\psi(x+\hat 1) \bar\psi(y)[1+\gamma_1]\psi(y-\hat 1)\Bigr\} \biggr\rangle\, .\label{d2thetadef}
\end{eqnarray}
The first derivative vanishes due to the parity invariance of the action. The second
one is finite and its magnitude strongly depends on the ``physical'' spatial volume of the system. The right hand side of
figure \ref{allfigs} shows how this observable approaches zero when the lattice size grows at fixed bare parameters.

At the moment we are still in the process of testing whether these (or other) observables will allow us to 
define a line of constant physics accurately enough.

\section{Conclusions}
To recover a chirally invariant Gross-Neveu theory from a lattice model with Wilson fermions requires the 
careful tuning of four bare parameters. We have introduced and tested several observables that can be used to define
a ``line of constant physics''. First simulation results are encouraging, but whether some continuum extrapolated
quantity can be calculated accurately enough to justify the effort in this model remains to be investigated. 

\section*{Acknowledgment}
We thank Rainer Sommer, Janos Balog and Jean Zinn-Justin for fruitful discussions.

\bibliographystyle{JHEP}
\bibliography{./references}

\providecommand{\href}[2]{#2}\begingroup\raggedright\begin{thebibliography}{10}

\bibitem{Gross1974}
D.~J. Gross and A.~Neveu,  {\em Phys. Rev.} {\bf D10} (1974) 3235.

\bibitem{Aoki1985}
S.~Aoki and K.~Higashijima,  {\em Prog. Theor. Phys.} {\bf 76} (1986) 521.

\bibitem{Knechtli2004}
F.~Knechtli, T.~Korzec, B.~Leder, and U.~Wolff,  {\em Nucl. Phys. B (Proc. Suppl.)} {\bf 140} (2004)
  785--787, [\href{http://xxx.lanl.gov/abs/hep-lat/0410018}{{\tt
  hep-lat/0410018}}].

\bibitem{Luperini1991}
C.~Luperini and P.~Rossi, {\em Ann. Phys.} {\bf 212} (1991)
  371--401.

\bibitem{Kenna:2001fs}
R.~Kenna and J.~C. Sexton,  {\em Phys. Rev.} {\bf D65} (2002) 014507,
  [\href{http://xxx.lanl.gov/abs/hep-lat/0103014}{{\tt hep-lat/0103014}}].

\bibitem{Karowski:1980kq}
M.~Karowski and H.~J. Thun, {\em Nucl. Phys.} {\bf B190} (1981) 61.

\bibitem{Forgacs1991}
P.~Forgacs, S.~Naik, and F.~Niedermayer, {\em Phys. Lett.} {\bf B283} (1992) 282--286.

\bibitem{Cohen1981}
Y.~Cohen, S.~Elitzur, and E.~Rabinovici, {\em Phys. Lett.} {\bf B104} (1981) 289.

\bibitem{Campostrini1987}
M.~Campostrini, G.~Curci, and P.~Rossi,  {\em Nucl. Phys.} {\bf B314} (1989) 467--518.

\bibitem{Hands1992}
S.~Hands, A.~Kocic, and J.~B. Kogut, {\em Ann. Phys.} {\bf 224} (1993) 29--89,
  [\href{http://xxx.lanl.gov/abs/hep-lat/9208022}{{\tt hep-lat/9208022}}].

\bibitem{Bochicchio:1985xa}
M.~Bochicchio, L.~Maiani, G.~Martinelli, G.~C. Rossi, and M.~Testa, {\em Nucl. Phys.} {\bf B262}
  (1985) 331.

\bibitem{Duane1987}
S.~Duane, A.~D. Kennedy, B.~J. Pendleton, and D.~Roweth,  {\em Phys. Lett.} {\bf B195} (1987) 216--222.

\bibitem{Luscher:1992an}
M.~L\"uscher, R.~Narayanan, P.~Weisz, and U.~Wolff, {\em
  Nucl. Phys.} {\bf B384} (1992) 168--228,
  [\href{http://xxx.lanl.gov/abs/hep-lat/9207009}{{\tt hep-lat/9207009}}].

\bibitem{Sint:1993un}
S.~Sint, {\em Nucl. Phys.} {\bf
  B421} (1994) 135--158, [\href{http://xxx.lanl.gov/abs/hep-lat/9312079}{{\tt
  hep-lat/9312079}}].

\bibitem{Leder2005}
B.~Leder and T.~Korzec, {\em PoS(LAT2005)266}.

\bibitem{Furuya:1982fh}
K.~Furuya, R.~E. Gamboa~Saravi, and F.~A. Schaposnik,  {\em
  Nucl. Phys.} {\bf B208} (1982) 159.

\end{thebibliography}\endgroup

\end{document}